\documentclass[aps,twocolumn,prl,superscriptaddress,showpacs,tightenlines]{revtex4}
\usepackage{amssymb}
\usepackage{amsmath}
\usepackage{graphicx}
\usepackage{epsfig}
\usepackage{subfigure}
\usepackage{amsfonts}

\begin{document}

\title{Phase gate of one superconducting qubit simultaneously controlling
$n$ qubits in a cavity}
\author{Chui-Ping Yang}
\affiliation{Advanced Science Institute, The Institute of Physical
and Chemical Research (RIKEN), Wako-Shi, Saitama 351-0198, Japan}
\address{Physics Department, The University of Michigan,
Ann Arbor, Michigan 48109-1040, USA}
\author{Yu-xi Liu}
\affiliation{Advanced Science Institute, The Institute of Physical
and Chemical Research (RIKEN), Wako-Shi, Saitama 351-0198, Japan}
\affiliation{Institute of microelectronics,Tsinghua University,
Beijing 100084, China}
\author{Franco Nori}
\affiliation{Advanced Science Institute, The Institute of Physical
and Chemical Research (RIKEN), Wako-Shi, Saitama 351-0198, Japan}
\address{Physics Department, The University of Michigan,
Ann Arbor, Michigan 48109-1040, USA}
\date{July 23, 2009}

\begin{abstract}
We propose how to realize a three-step controlled-phase gate of
one superconducting qubit simultaneously controlling $n$ qubits
selected from $N$ qubits in a cavity ($1<n<N$). The operation time
of this gate is independent of the number $n$ of qubits involved
in the gate operation. This phase gate controlling at once $n$
qubits is insensitive to the initial state of the cavity mode and
can be used to produce an analogous CNOT gate simultaneously
acting on $n$ qubits.
\end{abstract}

\maketitle
\date{\today}

{\it Introduction.}--- Quantum information processing has
attracted considerable interest during the past decade. The
building blocks of quantum computing are single-qubit and
two-qubit logic gates. So far, a large number of theoretical
proposals for realizing two-qubit gates in many physical systems
have been proposed.~Moreover, two-qubit controlled-not (CNOT) or
controlled-phase (CP) gates have been experimentally demonstrated
in, e.g., cavity QED~\cite{Turchette}, ion traps~\cite{Monroe},
NMR~\cite{Jones}, quantum dots~\cite{Li}, and superconducting
qubits~\cite{Yamamoto,Groot}.

Attention is now shifting to the physical realization of {\it
multi}-qubit controlled gates (e.g.,~\cite{Monz}) instead of just
{\it two}-qubit gates. It is known that multi-qubit controlled
gates play a significant role in constructing network quantum
computation circuits. When using the conventional
gate-decomposition protocols to construct a multi-qubit controlled
gate~\cite{Barenco,onen}, the procedure usually becomes
complicated (especially for large $n$), as the number of
single-qubit and two-qubit gates required for the gate
implementation heavily depends on the number $n$ of qubits.
Therefore, building a multi-qubit controlled gate may become very
difficult since each elementary gate requires turning on and off a
given Hamiltonian for a certain period of time, and each
additional basic gate adds experimental complications and the
possibility of more errors.

Several methods for constructing phase gates with $n$-control
qubits acting on one target qubit based on cavity QED or ion traps
have been recently proposed~\cite{Yang,Duan,Lin,Wang}. These
methods open a new way for realizing quantum controlled-phase
gates with multiple control qubits. However, we note that these
proposals~\cite{Yang,Duan,Lin,Wang} cannot be extended to perform
a different type of significant multi-qubit controlled-phase gate,
i.e., quantum controlled-phase gates with one qubit controlling
$n$ target qubits.
\begin{figure}[tbp]
\includegraphics[bb=80 385 560 521, width=8.6 cm, clip]{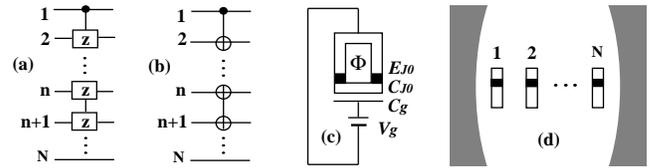}
\caption{(a) A controlled-phase (CP) gate simultaneously acting on
$n$ target qubits (2,~3,~...,~$n+1$), i.e., $n$ two-qubit CP
gates.~Here, Z represents a controlled-phase flip on each target
qubit.~Namely, if the control qubit (i.e., qubit 1) is in the
state $\left| 1\right\rangle $, then the state $\left|
1\right\rangle $ at each Z is phase-flipped as $\left|
1\right\rangle $ $\rightarrow -\left| 1\right\rangle $, while the
state $\left| 0\right\rangle $ remains unchanged.~(b) CNOT gate
simultaneously controlling $n$ qubits (2,~3,~...,~$n+1$), obtained
from our $n$-qubit CP gate.~The symbol $\oplus$ represents a CNOT
gate on each target qubit.~If the control qubit is in the state
$\left| 1\right\rangle $, then the state at $\oplus$ is bit
flipped as $\left| 1\right\rangle $ $\rightarrow \left|
0\right\rangle $ and $\left| 0\right\rangle $ $\rightarrow \left|
1\right\rangle $.~However,~when the control qubit is in the state
$\left| 0\right\rangle $,~the state at $\oplus$ remains unchanged.
(c)~Diagram of a superconducting (SC) charge qubit. (d)$N$ SC
qubits are placed in a microwave cavity, from which a subset of
qubits, selected for the gate, are coupled to each other via the
cavity mode.~In~(a)~and~(b),~the qubits ($n+2$,~$n+3$,~...,~$N$)
are not involved in the gate operation, by setting their
$\Phi=\Phi_0/2,~V_g^{\rm dc}=e/C_g,$~and $V_g^{\rm ac}=0$.}
\label{fig:1} \vspace*{-0.18in}
\end{figure}
\vspace*{-0.01in} In this work, we propose how to realize a
three-step controlled-phase gate of one superconducting (SC) qubit
simultaneously controlling $n$ qubits selected from $N$ qubits in
a cavity ($1<n<N$).~To achieve this, we construct an effective
Hamiltonian which contains interaction terms between the control
qubit and each subordinate or target qubit. We will denote this
$n$-target-qubits control-phase gate as a NTCP gate (see
Fig.~1(a)).~As shown below, our proposal has the following
advantages: (i) the $n$ two-qubit controlled-phase gates involved
in the NTCP gate can be performed simultaneously; (ii) the
operation time required for the gate implementation is independent
of the number $n$ of qubits involved in the gate operation; (iii)
this approach is insensitive to the initial state of the cavity
mode, and thus no preparation for the initial state of the cavity
mode is needed; (iv) no large detuning between the qubits and the
cavity mode is required and thus the gate operation can be speeded
up; and (v) the proposal is remarkably simple, requiring only
three basic operations. Note that a CNOT gate of one qubit
simultaneously controlling $n$ qubits, shown in Fig.~1(b), can
also be achieved using the present proposal. This is because the
$n$-target-qubits CNOT gate is equivalent to a NTCP gate plus a
single-qubit Hadamard gate acting on each target qubit before and
after the NTCP gate. To the best of our knowledge, our proposal is
the first to demonstrate that a powerful phase gate, {\it
synchronously controlling $n$ qubits}, can be achieved with
superconducting charge qubits in a cavity, which can be initially
in an arbitrary state.~This proposal is quite general and can be
easily extended to other physical systems (such as atomic qubits,
quantum dots, and superconducting flux or phase qubits), since the
relevant effective Hamiltonian [Eq.~(\ref{lps-10})~below] can be
constructed by applying suitable external driving pulses.~We
believe that this work is of general interest and significance
because it provides a simple protocol for performing
controlled-phase (or controlled-not) gates with
multiple-target-qubits, which are important in quantum information
processing such as entanglement preparation~\cite{sasura}, error
correction~\cite{Nielsen}, quantum algorithms (e.g., the Discrete
Cosine Transform~\cite{Rotteler}), and quantum
cloning~\cite{Braunstein}.

\vspace*{-0.01in}{\it Model.}--- The superconducting charge qubit
considered here, shown in Fig.~1(c),~consists of a small box,
connected to a symmetric superconducting quantum interference
device (SQUID) with capacitance~$C_{J0}$~and Josephson coupling
energy $E_{J0}.$ In the charge regime $\Delta~\gg~E_c\gg~E_{J0}\gg
~k_BT$ (here,~$k_B,$~$\Delta,$~$E_c,$~and~$T$ are the Boltzmann
constant, gap, charging energy, and temperature, respectively),
only two charge states, $n=0$ and $n=1$, are important for the
dynamics of the system, and thus this device~\cite{Makhlin}
behaves as a two-level system $\left\{ \left| 0\right\rangle
,\left| 1\right\rangle \right\}$.~For $N$ identical charge qubits
placed in a single-mode cavity~\cite{JQYou,Blais1} (Fig.~1(d)),
one can select a subset of qubits for the gate, while the
remaining qubits are not involved in the gate operation, by
setting their $\Phi=\Phi_0/2,~V_g^{\rm dc}=e/C_g,$~and $V_g^{\rm
ac}=0$ to have them decoupled from the cavity mode and their free
Hamiltonian being zero. Here, $\Phi _0$ is the flux quantum, $\Phi
$ is the external magnetic flux piercing the SQUID loop, $V_g^{\rm
dc}$ is the dc gate voltage, $V_g^{\rm ac}$ is the ac gate
voltage, and $C_g$ is the gate capacitance.~The method presented
below for a NTCP gate works for a subset of qubits selected from
the $N$ qubits in a cavity, because the qubit-qubit coupling,
mediated by the cavity mode, does not depend on the relative
position between any two qubits.~Without loss of generality, we
assume that the set of qubits involved in the gate operation are
the $n+1$ qubits labelled by 1,~2,~...,~and~$n+1$ (here, $1<n<N$).
The~Hamiltonian for the $n+1$ qubits and the cavity mode is
\vspace*{-0.01in}
\begin{eqnarray}
H&=&\hbar \omega _c\,a^{\dag}a+E_z\!\left( V_g^{\rm dc}
\right)S_z-E_J\!\left( \Phi \right) S_x \nonumber  \\
&&+\hbar \Omega \cos \left( \omega t+\varphi \right) S_z+\hbar
g\left( a+a^{\dag}\right) S_z,\label{lps-01}
\end{eqnarray}
where $a$ is the photon annihilation operator of the cavity mode
with frequency $\omega _c$; $S_z$ and $S_x$ are the collective
operators of the qubits, given by $S_z=\sum_{j=1}^{n+1}\sigma
_{z,j}$ and $S_x=\sum_{j=1}^{n+1}\sigma _{x,j}$, with Pauli
operators $\sigma _{z,j}=\left| 0\right\rangle _j\left\langle
0\right| -\left| 1\right\rangle _j\left\langle 1\right| $ and
$\sigma _{x,j}=\left| 0\right\rangle _j\left\langle 1\right|
+\left| 1\right\rangle _j\left\langle 0\right| $ for qubit $j$;
$g$ is the coupling constant between the cavity mode and each
qubit; and $\Omega $ is the Rabi frequency. In addition, $E_z$
$=-2E_c(1-2n_g^{\rm dc})$ with the charge energy
$E_c=e^2/(2C_g+4C_{J0})$ and $n_g^{\rm dc}=C_gV_g^{\rm dc}/\left(
2e\right) .$ The effective Josephson coupling is given by
$E_J\left( \Phi \right) =2E_{J0}\cos \left( \pi \Phi /\Phi
_0\right) .$ The fourth term of Eq.~(1) comes from the $\rm ac$
gate voltage given by $V_g^{\rm ac}=V_0\cos \left( \omega
t+\varphi \right) $ while the last term arises from the quantum
part of the gate voltage given by $V_g^{\rm qu}=V_0^{\rm qu}\left(
a+a^{\dag}\right) $, which is caused by the electric field of the
cavity mode when a qubit is inside the cavity. Finally, the
coupling constant $g$ and the Rabi frequency $\Omega$ are given by
$g=$ $2E_cC_gV_0^{\rm qu}/\left( \hbar e\right) $ and $\Omega
=2E_cC_gV_0/\left( \hbar e\right) ,$ respectively. Hereafter, we
set $E_z=0$ (i.e., $n_g^{\rm dc}=1/2$), $\omega _0=E_J\left( \Phi
\right) /\hbar $, and $\hbar =1$.~From the last two terms of
Eq.~(1), in the interaction picture with respect to $H_0=\omega
_ca^{\dag}a-\omega _0S_x$, we obtain (under a rotating-wave
approximation and assuming $\omega =2\omega _0$):
\begin{eqnarray}
H_1 &=&\frac \Omega 2\left[ S_z \cos \varphi \,+i
(S^{+}-S^{-})\sin \varphi\right] , \label{lps-02}\\
H_2 &=&\frac g2\left[ e^{-i\delta t}a\left(
S_z+S^{-}-S^{+}\right)+\text{H.c.}\right] , \label{lps-03}
\end{eqnarray}
where $S^{+}$ and $S^{-}$ are, respectively, the raising and
lowing operators for the qubits, given by
$S^{+}=\sum_{j=1}^{n+1}\left| 1\right\rangle _j\left\langle
0\right| $ and $S^{-}=\sum_{j=1}^{n+1}$ $\left| 0\right\rangle
_j\left\langle 1\right| ;$ and $\delta =\omega _c-\omega $ is the
detuning between the cavity-mode frequency and the frequency of
the ac gate voltage.

{\it Unitary evolution.}--- We now consider two
special cases: $\varphi =0$ and $\delta <0$, as well as $\varphi =\pi $ and $%
\delta >0.$ The results from the unitary evolution, obtained for
these two special cases, will be employed below for the gate
implementation. Let us begin with the case $\varphi =0$ and
$\delta <0.$ Performing the transformation $e^{-iH_1t}$ on $H_2$
for $\varphi =0$, and assuming $\Omega \gg \left| \delta \right|
,$ $g$, we obtain

\vspace*{-0.1in}
\begin{eqnarray}
H_2^{\prime }=e^{iH_1t}H_2e^{-iH_1t}=\frac g2\left( e^{-i\delta
t}a+e^{i\delta t}a^{\dag}\right) S_z. \label{lps-04}
\end{eqnarray}

The evolution operator $U^{\prime }$ for
$H_2^{\prime }$ takes the form~\cite{Wang,rensen,YDWang}
\begin{eqnarray}
U^{\prime }\left( t\right) =e^{-iA\left( t\right)
S_z^2}e^{-iB\left( t\right) S_za}e^{-iB^{*}\left( t\right)
S_za^{\dag}} \label{lps-05}
\end{eqnarray}
with $B(t)=ig(e^{-i\delta t}-1)/(2\delta)$ and
$A(t)=g[2B^{*}(t)-gt]/(4\delta).$ When $t=\tau =2\pi /\left|
\delta \right|,$ we have $B(\tau) =0$ and $A(\tau) =-g^2\tau
/(4\delta)$.~Then,~the evolution operator (in the Schr\"odinger
picture) of the qubit system is
\begin{eqnarray} U\left( \tau
\right) = e^{-iH_0\tau}e^{-iH_1\tau} U^{\prime
}\left( \tau \right) ~~~~~~~~~~~~~~~~~~~~~~~~~~~~~~\nonumber\\
= \exp(i\omega _0\tau S_x)\exp(-i\Omega \tau S_z/2)\exp(-i\lambda
\tau S_z^2), \label{lps-06}
\end{eqnarray}
where~$\lambda=-g^2/\left( 4\delta \right)>0.$~Note~that~for
$\omega _0\tau=m\pi$ (hereafter~$m$~is~an~interger),~one~has~$\exp
(i\omega _0\tau S_x)=\left( -1\right) ^m\prod _{j=1}^{n+1}I_j $,
where $I_j$ is the identity operator for qubit $j.$ Thus, the
operator $U$ reduces to
\begin{eqnarray}
U\left( \tau \right) =\exp(-i\Omega \tau S_z/2)\exp(-i\lambda \tau
S_z^2). \label{lps-07}
\end{eqnarray}

Let us now consider the case $\varphi =\pi $ and $\delta
>0.$ For convenience, we replace $\delta $ and $g$ by $\delta
^{\prime }$ and $g^{\prime }$, respectively. By setting
$E_z=E_J\left( \Phi \right) =\Omega =0$ for qubit $1$, the
coupling of qubit $1$ with the cavity mode becomes negligibly
small. The terms corresponding to $j=1$ can thus be dropped off
from $H,~H_0,~H_1,$ and $H_2.$ Note that the reduced
Hamiltonian~(\ref{lps-02})~and~(\ref{lps-03}) for $\varphi =\pi $
have forms similar to those for $\varphi =0$.~Therefore, it is
straightforward to show that under the condition $\Omega \gg
\delta ^{\prime },$ $g^{\prime },$ when $t=\tau ^{\prime }=2\pi
/\delta ^{\prime },$ the evolution operator of the qubits becomes
\begin{eqnarray} \widetilde{U}\left( \tau ^{\prime }\right)
=\exp(i\omega _0\tau ^{\prime }S_x^{\prime })\exp(i\Omega \tau
^{\prime }S_z^{\prime }/2)\exp(i\lambda ^{\prime }\tau ^{\prime
}S_z^{\prime 2}), \label{lps-08}
\end{eqnarray}
where $\lambda ^{\prime }={g^{\prime }}^2/\left( 4\delta ^{\prime }\right) ,$ $%
S_z^{\prime }=\sum_{j=2}^{n+1}\sigma _{z,j}$ and $S_x^{\prime
}=\sum_{j=2}^{n+1}\sigma _{x,j}.$ The operator $\widetilde{U}$ can
be reduced to
\begin{eqnarray}
\widetilde{U}\left( \tau ^{\prime }\right) =\exp (i\Omega \tau
^{\prime }S_z^{\prime }/2)\exp (i\lambda ^{\prime }\tau ^{\prime
}S_z^{\prime 2}) \label{lps-09}
\end{eqnarray}
for $\omega _0\tau ^{\prime }=m\pi $. Finally, we note that the
operator $U$ (or $\widetilde{U}$) does not include the photon
operator $a$ or $a^{+}$ of the cavity mode. Hence, the cavity mode
can be initially in an arbitrary state (e.g., in a vacuum state, a
Fock state, a coherent state, or even a thermal state).

{\it Implementation of a NTCP gate.}---~~The operations for the
gate implementation and the unitary evolutions after each step of
operation are listed below:

Step (i): Set $V_g^{\mathrm{dc}}=e/C_g$ and
$V_g^{\mathrm{ac}}=V_0\cos \left( \omega t\right) $ for each
qubit.~Choose $\Phi $ (applied to each qubit) and $\omega $
appropriately such that  $\omega _0(\Phi)=\omega /2=0.5m\omega
_c/\left( m-1\right) ,$ leading to $\delta =-\omega _c/\left(
m-1\right) <0$ ($m>1$) and thus $\omega _0\tau =m\pi $ for $%
\tau =-2\pi /\delta .$ One can see that this is the case discussed
above for $\varphi =0$ and $\delta <0.$ Thus, the $U$ of Eq.
(\ref{lps-07}) is the evolution operator for the qubit system for
the time $\tau =-2\pi /\delta$.

Step (ii): Set $V_g^{\mathrm{dc}}=e/C_g$, $V_g^{\mathrm{ac}}=0$,
and $\Phi
=\Phi _0/2$ for qubit $1;$ while set $V_g^{\mathrm{dc}}=e/C_g$ and $V_g^{%
\mathrm{ac}}=V_0\cos \left( \omega t+\pi \right) $ for qubits ($2,3,...,n+1$%
).~In addition, set $\omega _0(\Phi)=\omega /2=0.5\left(
m-2\right) \omega
_c/\left( m-1\right) $ for qubits ($2,3,...,n+1$), by choosing $\Phi $ and $%
\omega $ appropriately.~Accordingly, we have $\delta ^{\prime
}=\left| \delta \right|=\omega _c/\left( m-1\right) >0$ ($m>2$)
and thus $\omega _0\tau ^{\prime }=\left( m-2\right) \pi $ for
$\tau ^{\prime }=2\pi /\delta ^{\prime }.$ One can see that this
is the case discussed above for $\varphi =\pi $ and $\delta >0.$
Hence, for an evolution time $\tau ^{\prime }=2\pi /\delta
^{\prime }$, the evolution operator of the qubit system is
$\widetilde{U}$ in Eq.~(\ref{lps-09}).

When $\delta ^{\prime }=-\delta $ and $g^{\prime }=g$, we have
$\lambda ^{\prime }=\lambda $ and $\tau ^{\prime }=\tau $. In this
case, the joint time evolution operator, after the above two-step
operation, is
\begin{eqnarray}
U\left( 2\tau \right) &=& \widetilde{U}\left( \tau \right) U\left(
\tau \right)\nonumber\\
&=& \exp \left(-i \Omega \tau \sigma _{z,1}/2 \right) \exp \left(
-i2 \lambda \tau \sigma _{z,1}S_z^{\prime } \right). \nonumber
\end{eqnarray}
The condition $\delta ^{\prime }=-\delta $ is automatically
satisfied by the steps above.

Step (iii): Set $V_g^{\rm ac}=0$ and $\Phi =\Phi _0/2$ for each qubit. Set $%
V_g^{\rm dc}=2en_{g,1}^{\rm dc}/C_g$ for qubit 1 while $V_g^{\rm
dc}=2en_g^{\rm dc}/C_g$ for qubits ($2,3,...,n+1$). In addition,
adjust the cavity-mode frequency~\cite{Sandberg,Laloy} such that
it is highly detunned with the $\left| 0\right\rangle
\leftrightarrow \left| 1\right\rangle $ transition of each qubit.~
Thus, the Hamiltonian in Eq.~(1) for the qubit system becomes
$H=E_{z,1}\sigma _{z,1}+E_zS_z^{\prime },$ where
$E_{z,1}=-2E_c(1-2n_{g,1}^{\rm dc})$ while $E_z=-2E_c(1-2n_g^{\rm
dc}).$ For a time interval $\tau ,$ the corresponding evolution
operator is then given by
\begin{eqnarray}
\overline{U}\left( \tau \right) =\exp(-i E_{z,1}
 \tau \sigma _{z,1}/\hbar) \exp(-i E_z \tau S_z^{\prime }/\hbar). \nonumber
\end{eqnarray}

With~a~choice~of~$n_{g,1}^{\rm
dc}=0.5-\hbar(4n\lambda+\Omega)/(8E_c)$ and $n_g^{\rm
dc}=0.5-\hbar \lambda/(2E_c)$, one can find from $U(2\tau)$ and
$\overline{U}(\tau)$ above that the joint time evolution operator,
after the above three-step operation, is given by $U\left( 3\tau
\right) =\overline{U}\left( \tau \right) U\left( 2\tau \right)
=\prod_{j=2}^{n+1}U_p\left( 1,j\right)$, with $U_p\left(
1,j\right) =\exp[i2\lambda \tau \left( \sigma _{z,1}+\sigma
_{z,j}-\sigma _{z,1}\sigma _{z,j}\right)].$ For the qubit pair
($1,j$), we have $U_p( 1,j) \left| r_1\right\rangle \left|
s_j\right\rangle = \left| r_1\right\rangle \left| s_j\right\rangle
$ (with~$rs=00,~01,$~or~$10$),~while $U_p\left( 1,j\right) \left|
1_1\right\rangle \left| 1_j\right\rangle = \exp(-i8\lambda \tau)
\left| 1_1\right\rangle \left| 1_j\right\rangle ,$ where an
overall phase factor $\exp(i2\lambda \tau)$ is omitted.
This result~shows~that~for~$8\lambda \tau =\left( 2k+1\right) \pi $, i.e.,~$%
\left| \delta \right| =2g/\sqrt{2k+1}$ ($k$ is an integer), a
two-qubit controlled-phase gate described by $U_p\left( 1,j\right)
=I_{1j}-2\left| 1_11_j\right\rangle \left\langle 1_11_j\right|$
is achieved for the qubit pair ($1,j$). Here and below, qubit $1$
acts as a control while qubit $j$ as a target, and $I_{1j}$ is the
identity operator for the two qubits 1 and $j$. All above
conditions on $\delta$ can be satisfied with an appropriate choice
of $m$,~$k$,~$\omega_c$~and~$g$.

Finally,~we~have $U(3\tau)=\prod_{j=2}^{n+1}(I_{1j}-2\left| 1_11_j
\right\rangle \left\langle 1_11_j\right|),$ which demonstrates
that $n$ two-qubit controlled-phase gates are simultaneously
performed on the qubit pairs ($1,2$), ($1,3$),..., and ($1,n+1$),
respectively. Note that each qubit pair contains the {\it same}
control qubit (i.e., qubit $1$) and a {\it different} target
qubit. Hence, a NTCP gate with $n$ target qubits ($2,3,...,n+1$)
and one control qubit (i.e., qubit 1) is obtained after the above
three-step operation. A brief overview on this NTCP gate is in
Fig.~1(a).

It can be found from $U(3\tau)$ and $U_p(1,j)$ above that the
present method is based on an effective Hamiltonian
\begin{eqnarray}
H_{\rm eff}={\small \sum_{j=2}^{n+1}}H_{1j}=-2\lambda
\sum_{j=2}^{n+1}\left( \sigma _{z,1}+\sigma _{z,j}-\sigma
_{z,1}\sigma _{z,j}\right) , \label{lps-10}
\end{eqnarray}
which contains the interaction terms between the control qubit and
each target qubit, but does not include interaction terms between
any two target qubits.~Note that each term $H_{1j}$ in
Eq.~(\ref{lps-10}) acts on a different target qubit, with the same
controlled qubit, and that any two terms $H_{1j}$ for different
$j$ commute with each other. Therefore, the $n$ two-qubit
controlled-phase gates forming the NTCP gate can be simultaneously
performed on the qubit pairs ($1,2$), ($1,3$),..., and ($1,n+1$).

{\it Proposed experiment.}--- For this method to work, the
following conditions should be met. For steps (i) and (ii), the
Rabi frequency $\Omega $ needs to be much larger than
$g,~g^{\prime },~-\delta ,~\delta ^{\prime };$ the deviation from
the degeneracy point is given by $\varepsilon _0=\hbar \left(
\Omega +g\right)/\left( 4E_c\right) ,$ which needs to be a small
number to have the qubits working near the degeneracy point.~For
step (iii), from $n_{g,1}^{\rm dc}$ and $n_g^{\rm dc}$ above, we
obtain $\varepsilon_1=\hbar(4n\lambda+\Omega)/(8E_c)$ and
$\varepsilon _2=\hbar \lambda/(2E_c)$. Here, $\varepsilon _1$ is
the deviation from the degeneracy point for the control qubit 1
while $\varepsilon_2$ for the $n$ target qubits, which should be
sufficiently small to ensure that the qubits work near their
degeneracy points.

\begin{figure}[tbp]
\includegraphics[bb=53 454 528 576, width=8.65 cm, clip]{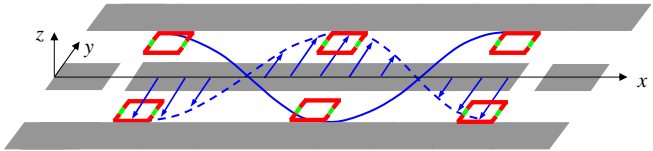}
\caption{(Color online) Proposed setup for six charge qubits (red
squares) and a (grey) standing-wave quasi-one dimensional coplanar
waveguide cavity. Each qubit is placed at an antinode of the
electric field. The two blue curves represent the standing-wave
electric field, along the $y$-direction.}
\label{fig:2}\vspace*{-0.05in}
\end{figure}

As a concrete example, let us consider the experimental
feasibility of implementing a five-target-qubit controlled-phase
gate using superconducting charge qubits with $C_g=1$ aF,
$C_{J0}=300$ aF, $E_c/h=32$ GHz, $E_{J0}/h=5$ GHz, $T_2=500$ ns,
and $T_1=7.3$ $\mu $s.~The charge qubits with these parameters are
available at present~\cite{Astafiev,Blais2}. For a superconducting
1D standing-wave coplanar waveguide cavity and each qubit placed
at an antinode of the cavity field (as shown in Fig.2), the
amplitude of the quantum part of the gate voltage is given by~\cite{Blais1} $%
V_0^{\rm qu}=\left( \hbar \omega _c\right) ^{1/2}\left(
Lc_0\right) ^{-1/2}$, where $L$ is the cavity length and $c_0$ is
the capacitance per unit
length of the cavity.~The coupling constant is then given by $%
g=2E_cC_g\left( \hbar e\right) ^{-1}\left( \hbar \omega _c\right)
^{1/2}\left( Lc_0\right) ^{-1/2},$ showing that $g$ does not
depend on the detuning $\delta $.~Therefore, the condition
$g=g^{\prime }$ required above can be satisfied. For charge qubits
with the
above parameters and a cavity with $\omega _c/\left( 2\pi \right) =10$ GHz, $%
L=\lambda \sim 12$ mm, $c_0\sim 0.22$ aF/$\mu $m, and $\varepsilon
_e=6.3,$~a simple calculation gives $g/2\pi \sim 100$ MHz, which
is experimentally available~\cite{Blais2}.~Here,~$\lambda $ is the
wavelength of the cavity mode and $\varepsilon _e$ is the
effective relative dielectric constant.~With the choice of $\left|
\delta \right| \sim 0.9g$ (corresponding to the integers $m=112,$
$k=2$), the total operation
time $t_{\rm op}=3\tau $ would be $%
\sim 33$ ns, which is much shorter than the dephasing time $T_2$
and the cavity-mode lifetime $\kappa ^{-1}=Q/\omega _c\sim 159$ ns
for a cavity with $Q=10^4$. Here, $Q$ is the (loaded) quality
factor of the cavity. Note that a quality factor $Q=10^4$ has been
demonstrated by cavity QED experiments with superconducting charge
qubits~\cite{Wallraff}.~For a qubit-cavity system with the
parameters given above, we have $%
\varepsilon _0\sim 5.46\times 10^{-3},$ $\varepsilon _1\sim
4.51\times 10^{-3},$ and $\varepsilon _2\sim 4.34\times 10^{-4}$
for $\Omega /\left( 2\pi \right) \sim 600$ MHz.~Therefore, the
conditions for the qubits to work near the degeneracy point are
well satisfied.

Note also that the adjustment of the cavity-mode frequency is
unnecessary because one can adjust the dc gate voltage to have the
qubits decoupled from the cavity mode. However, when a large
number of qubits are involved, it is much more convenient to
adjust the cavity mode frequency rather than adjusting the dc gate
voltage of each qubit. This is because all qubits can be
simultaneously decoupled from the cavity mode by adjusting the
cavity mode frequency, but one will need to individually adjust
step-by-step the dc gate voltage for each qubit to decouple the
qubits from the cavity mode.

We acknowledge partial support from the National Security Agency
(NSA), Laboratory for Physical Sciences (LPS), US Army Research
Office (ARO), National Science Foundation (NSF) under Grant No.
EIA-0130383, and the JSPS-RFBR under Contract No. 06-02-91200.

\end{document}